\documentclass[aps,prb,twocolumn,superscriptaddress,showpacs]{revtex4-1}
\usepackage{hyperref}
\usepackage{graphicx}
\usepackage{verbatim}
\include{commands}
\newcommand{\BFA}{BaFe$_2$As$_2$}
\newcommand{\dxz}{$d_{xz}$}
\newcommand{\dyz}{$d_{yz}$}
\newcommand{\kF}{k$_F$}

\newcommand{\ub}{$\mu_B$}

\newcommand{\FePn}{Fe$Pn_4$}
\newcommand{\LFAO}{LaFeAsO}
\newcommand{\OptFeCo}{Ba(Fe$_{0.92}$Co$_{0.08}$)$_2$As$_2$}
\newcommand{\FeCo}{Ba(Fe$_{x}$Co$_{1-x}$)$_2$As$_2$}
\newcommand{\CFA}{CaFe$_2$As$_2$}

\begin{document}


\title{Phonon Softening near structural transition in BaFe$_2$As$_2$ observed by inelastic x-ray scattering}


\author{Jennifer L. Niedziela}
\email[]{jniedzie@utk.edu}
\affiliation{Department of Physics and Astronomy, University of Tennessee - Knoxville, Knoxville, TN}
\affiliation{Oak Ridge National Laboratory, Oak Ridge, TN}

\author{D. Parshall}
\altaffiliation{Present Address Department of Physics, University of Colorado, Boulder, CO, 80302.}
\affiliation{Department of Physics and Astronomy, University of Tennessee - Knoxville, Knoxville, TN}

\author{K. A. Lokshin}
\affiliation{Department of Materials Science and Engineering, University of Tennessee - Knoxville, Knoxville, TN}

\author{A. S. Sefat}
\affiliation{Oak Ridge National Laboratory, Oak Ridge, TN}

\author{A. Alatas}
\affiliation{Advanced Photon Source, Argonne National Laboratory, Darien, IL}
\author{T. Egami}
\affiliation{Department of Physics and Astronomy, University of Tennessee - Knoxville, Knoxville, TN}
\affiliation{Oak Ridge National Laboratory, Oak Ridge, TN}
\affiliation{Department of Materials Science and Engineering, University of Tennessee - Knoxville, Knoxville, TN}


\date{\today}

\begin{abstract}
In this work we present the results of an inelastic x-ray scattering experiment detailing the behavior of the transverse acoustic [110] phonon in \BFA\ as a function of temperature. When cooling through the structural transition temperature, the transverse acoustic phonon energy is reduced from the value at room temperature, reaching a maximum shift near inelastic momentum transfer $\vec{q}=0.1$. This softening of the lattice results in a change of the symmetry from tetragonal to orthorhombic at the same temperature as the transition to long-range antiferromagnetic order. While the lattice distortion is minor, the anisotropy in the magnetic exchange constants in pnictide parent compounds is large.  We suggest mechanisms of electron-phonon coupling to describe the interaction between the lattice softening and onset of magnetic ordering.

\end{abstract}

\pacs{63.20.-e, 74.25.kc, 63.20.kd}

\maketitle


\section{Introduction}
Discovery of superconductivity in iron pnictide materials\cite{Hosono:2008} spurred intense experimental and theoretical effort to unravel the underlying physics of these compounds.  The presence of \FePn\ tetrahedra is universal in pnictide compounds, where iron atoms are arranged in a square planar lattice and tetrahedrally coordinated by pnictogen atoms. A common dynamic feature in pnictide families is a structural phase transition concomitant or in close vicinity with a transition to long range antiferromagnetic (AFM) order upon cooling\cite{DelaCruz:2008p50,Rotter:2008p28}. This transition is suppressed with hole\cite{Takahashi:2008p857,Chen:2008p5004,Rotter:2008p14},electron\cite{Sefat:2008p853,Sefat:2008p5575}, or isoelectronic doping\cite{Ren:2009p6790,Jiang:2009p6694}, eventually giving way to superconductivity.

At room temperature, the pnictide parent compound \BFA\ is a tetragonal ($I4/mmm$) paramagnet.  At T $\approx$ 140 K\cite{Rotter:2008p28}, \BFA\ undergoes a structural transition to orthorhombic ($Fmmm$) symmetry, and magnetically orders with spins AFM along the $a$-axis in the iron plane, ferromagnetic along the $b$-axis, and AFM between the iron layers with an ordered moment of 0.87(3) \ub\ per Fe site\cite{Huang:2008p44}. The measured moment is lower than that predicted by DFT calculations\cite{Lumsden:2010p7848}. While the structural distortion in the pnictides is small, the magnetic exchange anisotropy is very large when the exchange constants are calculated from an anisotropic Heisenberg model calculation which includes damping with very small interlayer magnetic coupling\cite{Zhao:2009p1337, Harriger:2010p9607}. 

For AFM ordering to occur, the high temperature tetragonal symmetry of \BFA\ must be broken to allow for the spin anisotropy, making the lattice distortion from tetragonal to orthorhombic symmetry a requirement for AFM ordering. The precise mechanism of the lattice distortion is a matter of considerable debate \cite{Kontani:2011p11703,Goto:2011p11704,Paul:2011p11708,Johannes:2010p10214, Johannes:2009p6015,Yildirim:2009p6935,Lee:2009p9621, Kontani:2010p11428}.

Early theoretical work shows that the lattice distortion\cite{DelaCruz:2008p50}  in the oxyarsenide \LFAO\ is the manifestation of relieved magnetic frustrations as the AFM ordering results in different occupancy for the \dxz, \dyz\ orbitals which breaks the tetragonal symmetry\cite{Yildirim:2009p6935}. However, the exchange constants determined from the spin wave dispersion are not consistent with this idea\cite{Zhao:2009p1337}. Another explanation for the coupling of the transitional behaviors is that of ferro-orbital ordering. A minimal approach applying the Hartree-Fock approximation to a two orbital model shows that AFM ordering arises as a consequence of ferro-orbital ordering, which then results in a lattice distortion\cite{Kubo:2009p8976}. Three- and five-orbital models suggest that orbital degrees of freedom are strongly coupled to magnetic ordering, with \dxz\ and \dyz\ susceptible to orbital ordering, but that the orbital magnetization is much greater for the \dyz\ orbital, resulting in a reduction of the orbital hybridization between the \dxz\ and \dyz\ orbitals\cite{Daghofer:2010p8688}.  Other reports suggest that orbital ordering is able to spontaneously break the $C_4$ symmetry locally without Fermi surface nesting or magnetic frustrations\cite{Lee:2009p9621}.  

Softening of the shear elastic constant in \FeCo\ has been noted in several works including resonant ultrasound spectroscopy on \BFA\ and \OptFeCo,\cite{Fernandes:2010p11227} and ultrasonic pulse echo measurements on undoped, underdoped\cite{Yoshizawa:2010p11294}, and optimally doped \FeCo \cite{Goto:2011p11704}.  The measurements showed a softening of the $C_{66}$ mode near the structural or superconducting transition in all compounds where measurement wasn't precluded by sample instability against the mounting apparatus. The $C_{66}$ mode is responsible for structural phase transitions of type which serve to lower crystalline symmetry from tetragonal to orthorhombic\cite{CowleyBook,Folk:1976p11701}, so this observation is expected, although in the presented works several additional mechanisms for the phase transition are proffered. 
   
Resonant ultrasound spectroscopy measurements performed on \BFA\ show a strong softening of the shear elastic constant, which is used to conclude that there is general lattice softening through the structural transition driven by renormalization of the shear elastic constant by nematic fluctuations\cite{Fernandes:2010p11227}. In this context, the softening of the lattice is a secondary effect, with electronic degrees of freedom taking precedence over elastic, and the structural transition then a mere consequence of the magnetic ordering.  In the work conducted in ultrasonic pulse echo spectroscopy, the softening minima tracks with the superconducting temperature under application of a magnetic field\cite{Yoshizawa:2010p11294, Goto:2011p11704}, and other elastic constants show almost imperceptible deviations from monotonic behavior in close proximity to the superconducting transition temperature\cite{Goto:2011p11704}.  One hypothesis derived from these experiments is the phase transition is driven by ferro-orbital ordering, suggestive of strong lattice coupling and orbital fluctuations\cite{Yoshizawa:2010p11294}, while another concludes that the softening is induced by the coupling of the degenerate electronic orbitals to the elastic strain\cite{Goto:2011p11704}.

There is some disagreement between the reports on the elastic constant behavior at low temperature.  Upon cooling below the structural or superconducting temperature, some reports show the value of $C_{66}$ recovering some 10\% of the room temperature value at low temperature for the undoped and superconducting samples\cite{Fernandes:2010p11227}, while other reports on the underdoped show no further observed softening below the structural transition temperature\cite{Goto:2011p11704}.

Experimental studies on the transverse acoustic (TA) modes of the related compound \CFA\ shows a softening of the transverse acoustic mode at the zone boundary for a TA [110] mode, and a mid-zone point at TA [100] at the magneto-structural transition when measuring accompanied by phonon broadening\cite{Mittal:2009p6713}. Further, calculation shows that phonons are very sensitive to the structural details of the \FePn\ layers, with phonon modes that distort the tetrahedra having substantial impact on the electronic density of states close to $E_F$\cite{Zbiri:2009p6304}.  Anomalous phonon behavior is observed in Raman spectroscopy, with the $E_g$ in-plane phonon mode split in \BFA\cite{Chauviere:2009p10235,Schafgans:2011p11709,Rahlenbeck:2009p11398}, which is linked to strong spin-phonon coupling. This $E_g$ mode is a doubly-degenerate mode impacting Fe and As, and the contention is that the split is too large to be accounted for by the orthorhombic distortion and instead is indicative of spin-phonon coupling.  Measurements of phonon behavior conducted with inelastic x-ray spectroscopy are replicated well by theory when magnetism is explicitly taken into account\cite{Reznik:2009p9602}, indicating the presence of spin-phonon coupling.

The degree of hybridization between the As-$4p$ and Fe-$3d$ states is shown to have strong impacts on the Fe$^{2+}$ spin state\cite{Yildirim:2009p46}, and to influence the Fe-As phonon modes\cite{Lee:2009p9621}.  Further, the dependence of the superconducting critical temperature\cite{Lee:CriticalAngle,Mizuguchi:2010p8333} and magnetic moment\cite{DeLaCruz:2010p6857,Egami:2010p8152} of pnictide materials as a function of Fe-As layer separation is well established.

The preceding discussion sets forth a clear need to understand the specifics of the structural transition as they relate to the elastic degrees, magnetic, and orbital degrees of freedom. Here we report the results of inelastic x-ray spectroscopy (IXS) measurements on single crystals of \BFA\ as a function of temperature. Since the acoustic phonon frequency goes as the square root of the elastic constant, we expect a softening of the low-energy acoustic phonon modes of \BFA\ when passing through the structural transition temperature. Phonon calculations predict a linear dispersion in the low-energy, long wavelength regime\cite{Zbiri:2009p6304,Reznik:2009p9602}. The acoustic phonon modes in \BFA\ can be characterized with energy transfers of less than 10 meV, and the small and constant momentum resolution offered by IXS allows precise measurements of the phonon dispersion as a function of temperature. The previous measurements on the shear elastic constant\cite{Fernandes:2010p11227} provide information only at the zone center ($\Gamma$ point, $\vec{q}$ = 0), so the IXS measurements described in this work provide additional information on the low-energy lattice dynamics in the \BFA\ system.

\section{Experiment}
We investigated the behavior of the transverse acoustic phonon mode in \BFA\ near the (2,2,0)$_{O}$ position, which is the [110] mode in the reduced scheme. The orthorhombic notation and the $Fmmm$ symmetry are used throughout this report. The measurements were made using the HERIX instrument on beamline 3-ID of the Advanced Photon Source at Argonne National Laboratory\cite{Toellner:2006p11270,Sinn:2001p11269}. The HERIX instrument is on an insertion device beamline with a six-bounce monochromator setup providing an incident energy of 21.657 keV, and a resolution of 2.3 meV at the elastic line.

Our sample was a single crystal of \BFA\ measuring 1mm x 2mm x 60 $\mu$m, grown using a self-flux method\cite{Sefat:2008p853}. Calculated optimal thickness for a transmission mode geometry for an incident energy of 21.657 keV was 55 $\mu$m. The crystal was oriented in transmission mode with (2, 2, 0) and (1, 1, 3) in the scattering plane. This alignment allows measurement of the phonons in the basal plane of the form $\vec{Q}=\vec{G}+\vec{q}=(\vec{H}-\vec{q},\vec{K}+\vec{q})$, where $\vec{G}=(\vec{H},\vec{K})$ is the reciprocal lattice vector of the Bragg peak in the $a-b$ plane, and $\vec{q}$ the inelastic momentum transfer. For temperature dependent studies, the crystal samples were mounted on copper holders with a small amount of varnish, and the copper holders were attached to the cold finger of a closed cycle cryostat sealed by a beryllium dome. 

The beam size on sample was 50 $\mu$m by 350 $\mu$m. Only one analyzer-detector pair was used for these measurements, and an analyzer mask was used to limit Bragg contamination for low $\vec{q}$ measurements.  The mask size was 4 cm by 4 cm at $\vec{q} \geq 0.07$, and 1 cm by 1 cm for all $\vec{q} < 0.07$. The Stokes and anti-Stokes profiles were measured in most cases, but at points where an anti-Stokes contribution to the phonon spectra was dynamically limited, a contribution from elastic scattering was ensured by transversely scanning the crystal space at zero energy transfer. All data were normalized to the incident monitor count, a low efficiency ion chamber. 

For all temperatures, the phonon spectra were fit with Voigt profiles, employing the Whiting approximation to determine the width\cite{Whiting:1968p8948,Olivero:1977p8900}. Phonon spectra were corrected for the Bose thermal population factors. 

Fitted phonon peak widths are constrained to that of a phonon measured at $\vec{q} = 0.2$ at T = 130 K. Data for values of $\vec{q} > 0.07$ were fit with two phonon peaks and an elastic contribution. The phonon energy for $\vec{q} > 0.07$ is determined by the average of the fitted peak positions of the Stokes and anti-Stokes contributions. 

\begin{figure}[t]
   \centering
   \includegraphics[scale=0.40]{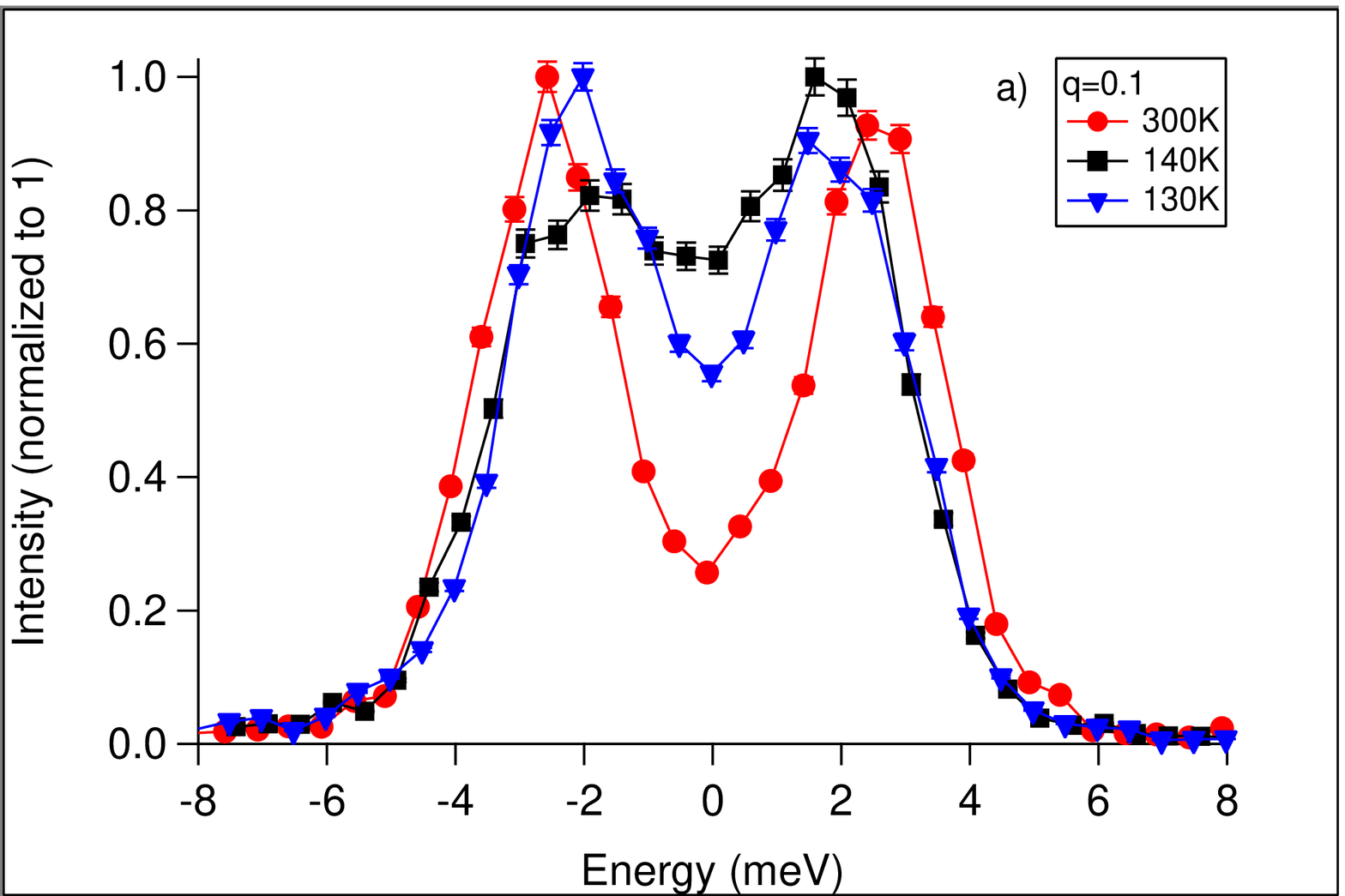} 
   \includegraphics[scale=0.40]{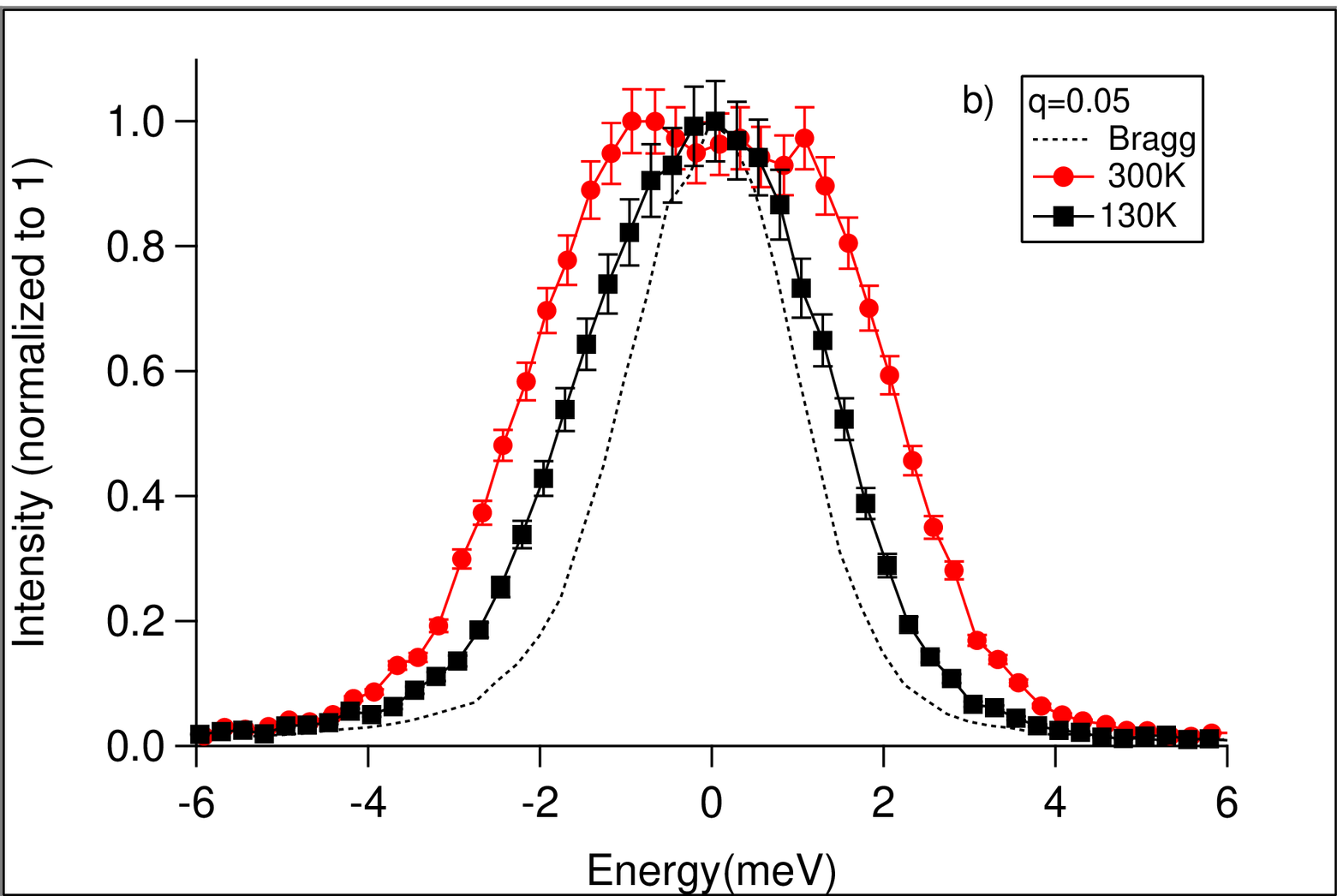} 
    \caption{(Color online) a) Comparison of 300 K, 140 K, and 130 K phonon spectra at $ \vec{q}=0.1$.  b) Comparison of 300 K and 130 K phonon spectra at $ \vec{q}=0.05$. The dotted line shows the Bragg peak measurement at T=130 K, $\vec{Q}$=(2,2,0)$_{O}$. All spectra are Bose factor corrected and normalized to one.  Errors are statistical errors propagated through Bose correction. }
    \label{fig:spectra} 

\end{figure}

Data for $\vec{q} \leq 0.07$ at temperatures below 300 K can not be resolved as distinct peaks, so the width of the observed peak is used as a proxy for the degree of softening (Fig. \ref{fig:spectra}). The peak width is determined using a single Voigt profile. 

\begin{figure}[t]
\includegraphics[scale=0.45]{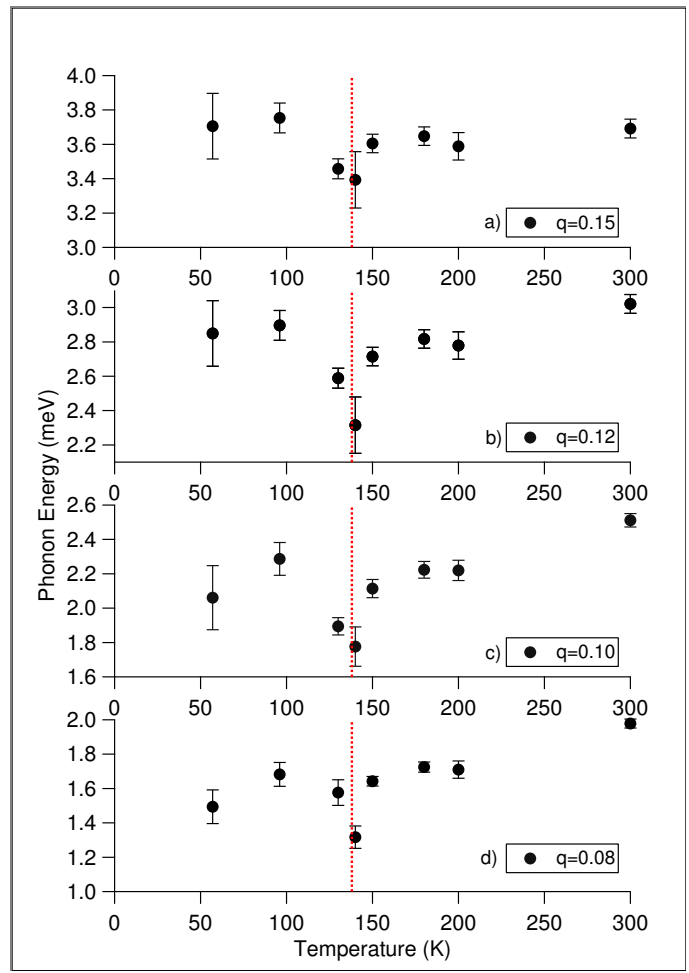}

\caption{(Color online) Energy of transverse acoustic phonon as a function of temperature for $\vec{Q}=(\vec{H}-\vec{q},\vec{K}+\vec{q})$ for $H=K=2$ and $\vec{q}=$ (a) 0.15, (b) 0.12, (c) 0.10, (d) 0.08 in orthorhombic reciprocal lattice units.  Dotted line indicates $T_S$.  Error bars are statistical errors propagated through Bose correction and peak averaging.  Two softenings are observed, one at $T_S$ and additional softening at low $\vec{q}$ at low temperature.}
\label{fig:dispersions_stack}
\end{figure}

\begin{figure}[htbp]
   \centering
   \includegraphics[scale=0.4]{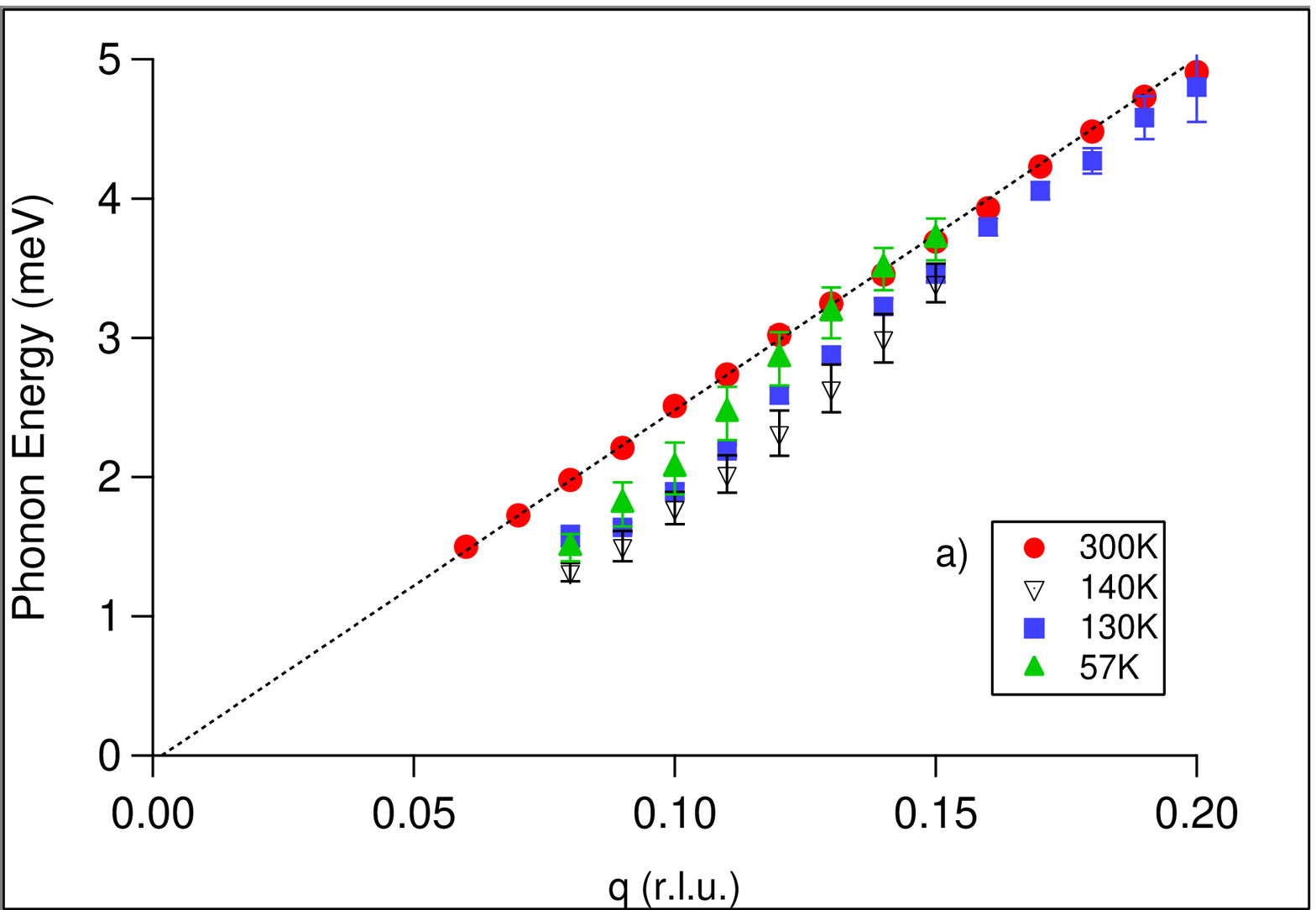}
      \includegraphics[scale=0.4]{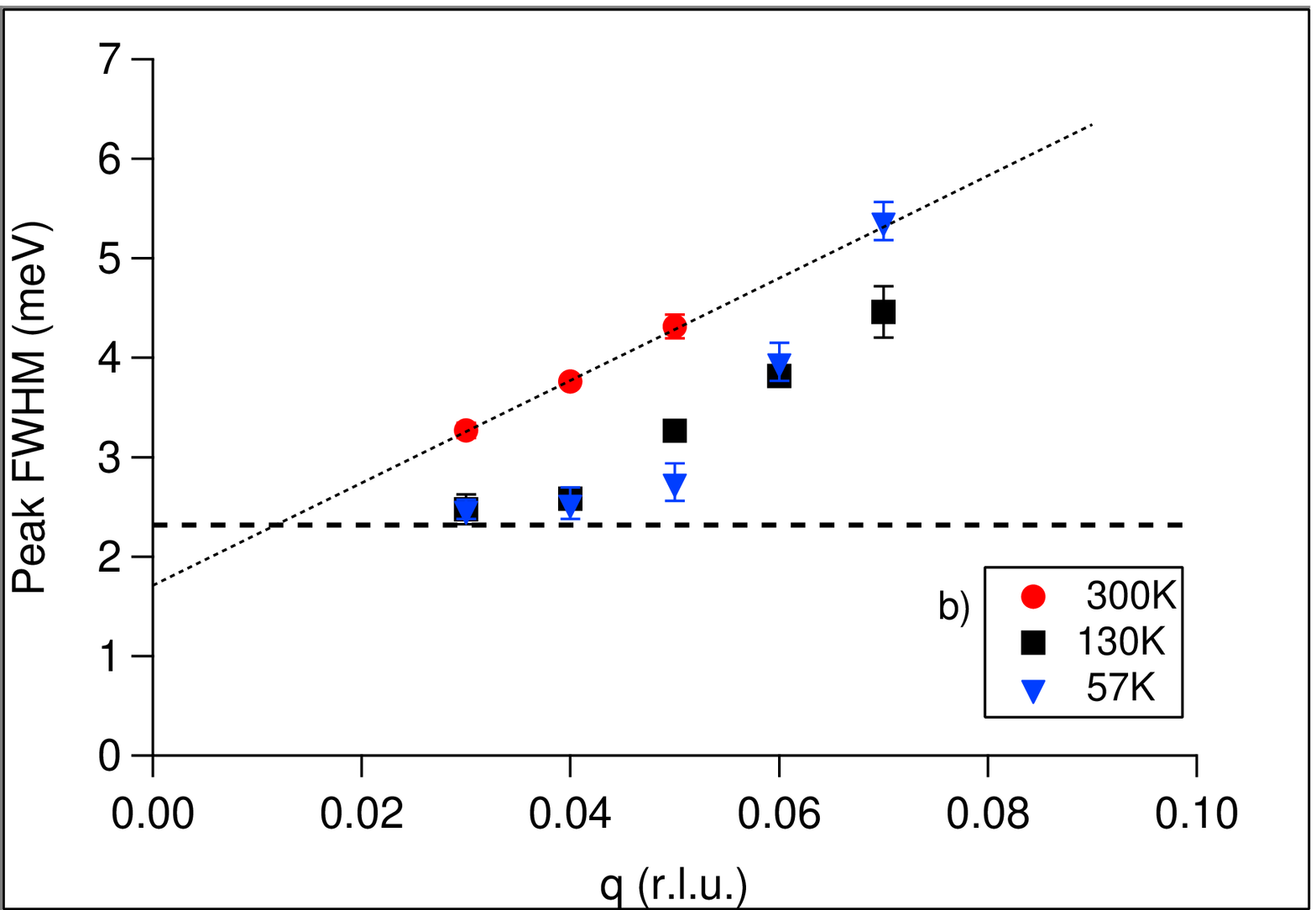} 
   \caption{(Color online) a) Dispersion of TA phonon at 300 K, 140 K, 130 K, and 57 K. Dotted line indicates a linear fit to the 300K data.  b) Central peak width as a function of $ \vec{q}$ for T=300 K, 130 K, and 57 K.    Heavy dashed line at $E=2.3$ meV indicates the width of the Bragg peak measured at $\vec{Q}=(2,2,0)$ T=130 K. Dotted line on 300 K data represents the fitted peak width.  Error bars are statistical errors propagated through Bose correction and the curve fitting procedure.}
   \label{fig:loq_vs_300}
\end{figure}

\section{Discussion}
The primary result of our investigation is the observation of a reduction in energy of the [110]$_O$ transverse acoustic phonon in the low $\vec{q}$ region, evident from the measured phonon spectra (Fig. \ref{fig:spectra}). The data show this reduction is most pronounced at $\vec{q} = 0.1$, reaching a maximal amount near T = 140 K (Fig. \ref{fig:dispersions_stack}). As the temperature is decreased, the phonon energy decreases, and then recovers, but not to the full energy observed at room temperature (Fig. \ref{fig:dispersions_stack}) before decreasing again at low temperatures and low $\vec{q}$. 

For data with $\vec{q} \leq 0.07$ the fitted width of the peak centered around zero energy transfer exhibits a linear behavior for 300 K (Fig. \ref{fig:loq_vs_300}a), with the width of the spectra collapsing into the Bragg peak width for lower temperatures (Fig. \ref{fig:loq_vs_300}b). 

The lattice softening observed in the shear elastic measurements show a drop in the shear mode to a minimal values at $T_S$, and then a slight recovery to low temperatures\cite{Fernandes:2010p11227,Goto:2011p11704}, or no recovery at all\cite{Yoshizawa:2010p11294}. This is in contrast to our observations here, as the phonon energy recovers most of the energy after softening, and the slope of the dispersion changes at low temperature. 

The only mode investigated in this work was the low energy transverse acoustic mode, and there was only a local softening of the mode - the traditional linear dispersion is recovered by $\vec{q} = 0.20$ in all temperature cases (four temperature cases shown in Fig. \ref{fig:loq_vs_300}).  This result is not captured in published DFT-GGA calculations reviewed in preparation of this manuscript. 

The softening of the low energy transverse acoustic mode observed by IXS is expected in the context of soft mode transitions, as the decrease in energy of the long wavelength phonons is always seen in structural phase transitions of the type observed here\cite{CowleyBook}.   But it has a few possible interpretations for microscopic mechanism. 

The first interpretation invokes Fermi surface nesting, which results in a regime of enhanced electron-phonon coupling. The Fermi surface structure of pnictides is well established to consist of hole pockets centered at the zone center, and electron pockets at the zone boundary\cite{Singh:2008p1908,Liu:2010p10089}, with electron or hole doping serving to change the levels of the two pockets relative to one another\cite{Sales:2010p10315}.  The softening of the TA phonon has a natural explanation in two-dimensional Fermi surface nesting giving rise to a Kohn anomaly, where strong quasiparticle excitations below $\vec{q}=2$\kF\ result in enhanced electron-phonon coupling \cite{Kohn:1959p6982}. Indeed we observe softening from the zone center to a maximal phonon softening below $\vec{q}$ = 0.1, which is consistent with the value of 2\kF\ = 0.1 for the inner hole pocket in \BFA\cite{Brouet:2009p5724}.  

In a work on \CFA, a series of line broadenings is observed\cite{Mittal:2009p6713}, though this is not used to draw the conclusion of Kohn anomalies, but may be related to the work here. However, possible broadening of the phonons for the \BFA\ compound studied here is too small to be fully resolved by the current technique.  Line broadenings of this type have been observed in studies of elemental superconductors, where the phonon anomalies were fully resolved only when using high resolution spin-echo neutron spectroscopy\cite{Aynajian:2008p6981}.  In order to confirm the presence of phonon broadening and/or a Kohn anomaly in \BFA, a careful analysis of the phonon lineshape using a higher energy resolution is required. Note also that there is some softening tendency at temperatures much below $T_s$. The $\vec{q}$ range of this low temperature softening is distinct from the softening at $T_s$; below $\vec{q} = 0.12$ at low temperatures, whereas softening is observed below $\vec{q} = 0.16$ near $T_s$. It appears that this softening has a different origin, and either one could be related to a Kohn anomaly.

A second interpretation is to consider the phonon softening in the context of coupling between the low energy acoustic mode and the orbital fluctuations.  In this scenario, softening of the shear mode causes in-plane shifting of the iron atoms, impacting the interatomic distances along the \FePn\ layer.  As these atoms shift their positions, the tetragonal orbital degeneracy between the \dxz\ and \dyz\ orbitals is broken, allowing the manifestation of magnetic order. 
While the orthorhombic structural distortion is small (of order 0.3\%, with the change in the iron square lattice 2.8 \AA\ on a side to a rectangle 2.79/2.81 \AA), and several works show that there is a strong dependence of the electronic structure on small details of the lattice, in particular the height of the pnictogen atom above the iron plane\cite{Singh:2008p1908, Singh:2008p840}.  Additionally, it has been shown that there is an enhancement of electron-boson coupling arising from orbital anisotropy\cite{Lee:2009p9621}, behavior which is in contrast to lower mean field values for the estimate of such couplings\cite{Boeri:2008p849,Boeri:2010p8222}.  Further, the pnictide systems are very close to the Stoner criticality, where even a small lattice distortion can have a large consequence to the magnetism in the system\cite{Egami:2010p8152}.  

The proposed mechanism indicates that other phonon modes may be involved in the structural transition via a similar softening behavior\cite{Egami:2010p8152}. Modes that directly influence the dynamics of the FeAs$_4$ tetrahedra are being examined.  Further, a coupling of acoustic and optical modes as is seen in other structural phase transitions\cite{Shapiro:2010p7768} is not ruled out, and requires further examination.

\section{Conclusions}
We have investigated the behavior of the transverse acoustic phonon in \BFA\ using IXS, and found that this phonon mode softens as a function of temperature, with a maximal softening near $\vec{q}$ = 0.1. The energy of the phonon mode drops well below that observed at room temperature, reaching a minimum near the structural phase transition temperature before recovering almost fully.  This observation is explained in terms of electron-phonon coupling resulting  in a lifting of the orbital degeneracy in the paramagnetic tetragonal state, allowing full manifestation of the AFM order.

\begin{acknowledgments}
This work was supported by the Office of Basic Sciences, US Department of Energy through the EPSCoR grant, DE-FG02-08ER46528 (JN, DP, KL and TE), and the Basic Energy Sciences, Materials Sciences and Engineering (AS). This work at the Advanced Photon source was supported by the U.S. Department of Energy, Office of Science, Office of Basic Energy Sciences, under Contract no DE-AC02-06CH11357.
\end{acknowledgments}


\end{document}